# SiZer for time series: A new approach to the analysis of trends

**Vitaliana Rondonotti**[*,§]

*Monetary Financial Institutions and Markets Statistics Division*
*European Central Bank, Frankfurt am Main, Germany*
*e-mail:* `vitaliana.rondonotti@ecb.int`

**J. S. Marron**[†,§]

*Department of Statistics and Operations Research*
*University of North Carolina, Chapel Hill, NC*
*e-mail:* `marron@email.unc.edu`

**Cheolwoo Park**[‡]

*Department of Statistics*
*University of Georgia, Athens, GA*
*e-mail:* `cpark@stat.uga.edu`

**Abstract:** Smoothing methods and SiZer are a useful statistical tool for discovering statistically significant structure in data. Based on scale space ideas originally developed in the computer vision literature, SiZer (SIgnificant ZERo crossing of the derivatives) is a graphical device to assess which observed features are 'really there' and which are just spurious sampling artifacts.

In this paper, we develop SiZer like ideas in time series analysis to address the important issue of significance of trends. This is not a straightforward extension, since one data set does not contain the information needed to distinguish 'trend' from 'dependence'. A new visualization is proposed, which shows the statistician the range of trade-offs that are available. Simulation and real data results illustrate the effectiveness of the method.

## 1. Introduction

Smoothing methods in statistics provide a useful tool for showing structure in data. Many monographs on smoothing are available in the statistical literature which in the last years include [2, 9, 11, 15, 22, 23].

[*]Rondonotti's work was supported in part by Università degli Studi di Roma 'La Sapienza', Dipartimento di Statistica, Probabilitá e Statistica Applicata.
[†]Marron's research was supported by NSF Grant DMS-9971649.
[‡]Park's research was supported by UGA Faculty Research Grants Program.
[§]This work is from the PhD dissertation of the first author, under direction of the second.

When smoothing methods are used for exploratory data analysis, a question that often arises is: which features are 'really there' (i.e. represent important underlying structure) and which are just spurious sampling artifacts (i.e. reflect unimportant random variation)? [5] proposed answering this question with a graphical device called SiZer (SIgnificant ZERo crossings of the derivatives), which is based on scale space ideas from computer vision.

Scale space is a family of kernel smooths indexed by the 'scale' $h$, which is usefully viewed as the level of resolution of the data. The scale $h$ is usually called the 'smoothing parameter', or the 'bandwidth' in the statistical literature. [19], for example, suggest using an overlay of these smooths for data analysis, and call it the 'family plot'. This method is illustrated in Figure 1 (a), using the Chocolate data set, i.e. the monthly production of chocolate in Australia from July of 1957 to October of 1990 (kilotonnes). This data set comes with the software companion to [4]. The dots show the Chocolate production after deseasonalising and linearly detrending the time series. The family of smooths suggests at different levels of resolution a dip around the year 1978 preceeded by two minor bumps and an increase in the last years. At the finest levels of resolution, many other features appear. Are all these features 'really there'? SiZer answers this question by assessing statistical significance of such features.

In particular, SiZer extends the usefulness of the family plot by visually displaying the statistical significance of features over both location $t$ and scale $h$. SiZer is based on confidence intervals for the derivative of the underlying function. The graphical device is a grey-scale map, reflecting statistical significance of the slope at $(t, h)$ locations in scale space. At each $(t, h)$ location, the curve is significantly increasing (decreasing) if the confidence interval is above (below) 0, so that map location is colored black (white). If the confidence interval contains 0, the curve at the level of resolution $h$ and at the point $t$ does not have a statistically significant slope, so the intermediate grey shade is used. Finally, if there is not enough information in the data set (according to a rule that will be illustrated in the following sections), at this scale space $(t, h)$ location, then no conclusion can be drawn, so the darker shade of grey is used to indicate that the data are too sparse.

The SiZer view of the Chocolate data is shown in Figure 1 (b). This data set was studied in [18], who viewed the errors, which will be defined later in (1), as independent and identically distributed (i.i.d.).

At the coarsest levels of resolution (largest bandwidths), at the top of the SiZer map, the intermediate grey appears everywhere, indicating that there is no statistically significant increase or decrease in the corresponding smooths. As it can be seen Figure 1 (a), these curves are close to the simple linear regression line, so the conclusion is that this line has no significant slope. Moving down in the plot, meaning decreasing the scale (i.e. the bandwidth gets smaller) the smooths are first significantly decreasing (white), then significantly increasing (black) after a short time interval in which the smooths are neither significantly increasing nor decreasing (intermediate grey). This shows that the minimum near year 1978 is statistically significant. As the bandwidth gets smaller, i.e. further down the plot, some additional significant features appear but only for a limited number of bandwidth values. At the finest level of resolution (smallest bandwidth) there is not enough information to assess the significance of any structure in the data set (darker shade of grey).

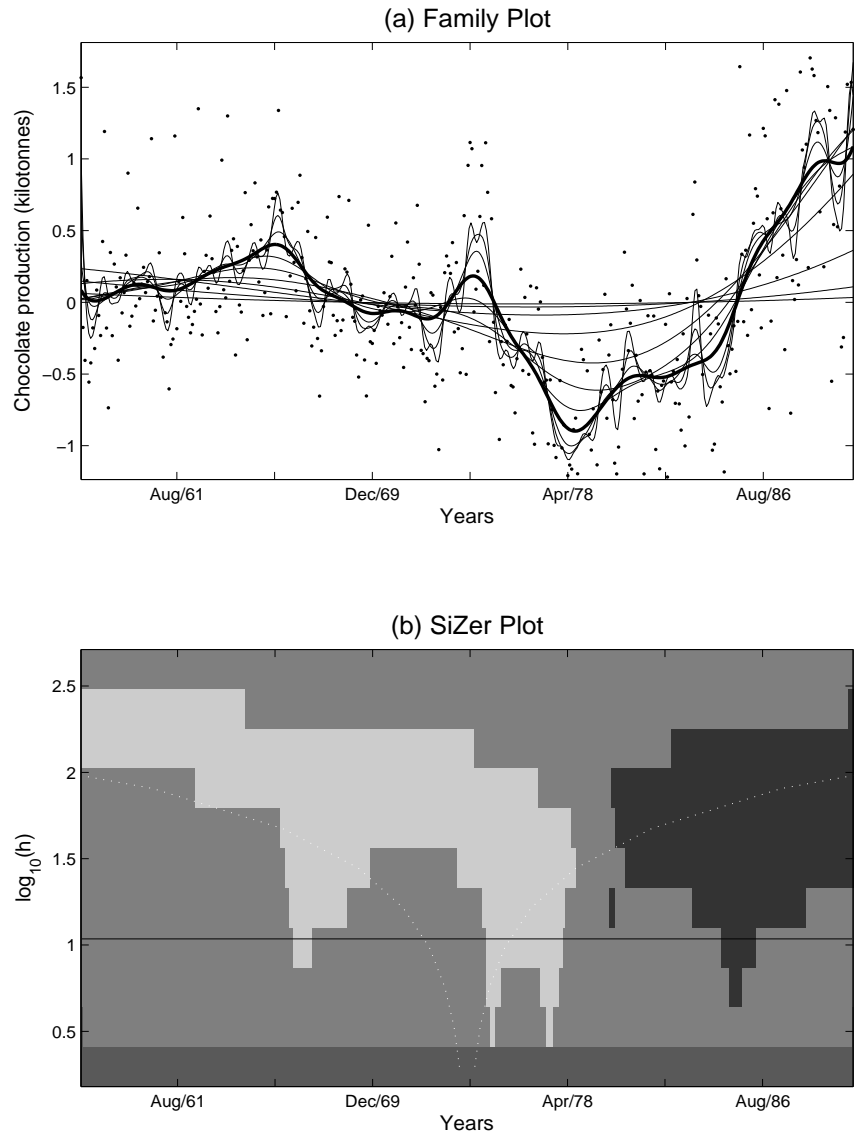

Fig 1. *Exploratory analysis of the Chocolate data set through the SiZer approach: (a) Family Plot; (b) SiZer Plot.*

Overall, the SiZer map reveals that only the decrease followed at around year 1978 by the increase in chocolate production are important features of the data.

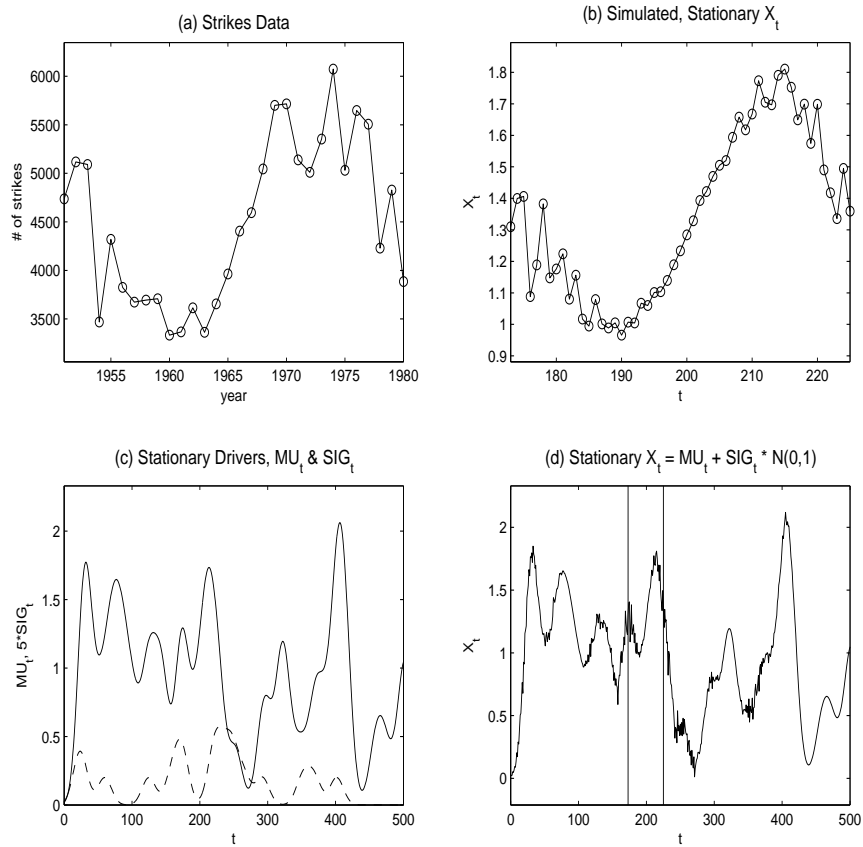

Fig 2. *An example for the identifiability problem between 'trend' and 'dependence artifacts': the Strikes data set.*

As pointed out by [5], the statistical inference, which is the basis of SiZer, makes heavy use of the assumption of i.i.d. errors. This assumption is inappropriate in time series contexts, where dependence is omnipresent, and in fact is usually the focus of statistical analysis. For SiZer to fulfill its potential to flag significant trends in time series, its underlying confidence intervals must be adjusted to properly account for the correlation structure of the data. This adjustment is straightforward when the correlation structure is known. But in the more important and common case of unknown correlation, SiZer for time series is not a straightforward extension. This is because of the identifiability problem between 'trend' and 'dependence artifacts'. Distinction between these can never be made on the basis of a single time series, as shown in Figure 2.

Figure 2 (a) shows the Strikes data from [4]. One view of this series is that its underlying distribution is nonstationary. In particular the mean function seems to decrease at the beginning, with an increase in the middle, and perhaps another decrease later. The variance also appears to be time varying, being large at both ends, and smaller in the middle. But this apparent 'nonstationarity' can not be proven on the basis of that single time series, as there are stationary stochastic processes whose realizations can look quite similar. An example is the simulated series shown in Figure 2 (b), which is a single realization of a stationary stochastic process. Note that the qualitative features, in terms of mean and variance, are very similar to those of Figure 2 (a). The remaining parts of Figure 2 reveal how this simulation was performed. The particular series of Figure 2 (b) was carefully selected from the much longer series shown in Figure 2 (d). It is the part between the vertical bars, chosen to match qualitatively the features of Figure 2 (a). The long series in Figure 2 (d) was generated as $X_t = \mu_t + \sigma_t Z_t$ where the curves $\mu_t$ and $\sigma_t$ are generated in the same way as drivers of the intensity function of a Cox process (shown in Figure 2 (c)), and where the $Z_t$ are independent standard normal realizations. Because $\mu_t$ and $\sigma_t$ are stationary stochastic processes, so is $X_t$.

Now consider this example from the viewpoint of 'trend'. The time series in Figures 2 (a) and (b) can be viewed as containing a large amount of 'trend' in the mean, but that structure can be equally well explained as 'serial dependence artifacts', or in fact some of each. The challenge for applications of SiZer in the trend estimation context is that these cases cannot be distinguished. This paper proposes an approach to this dilemma via a visualization which displays the range of trade-offs to the statistician.

Background on the local linear fit and on the SiZer approach will be found in Section 2. In Section 3 we will discuss the development of SiZer for studying trends in time series. The performance of our method is studied via simulated and real data sets in Sections 4 and 5.

## 2. Background

The application of nonparametric kernel techniques to estimating a deterministic trend has been popular in time series analysis. [1, 6, 14] used the kernel smoother for trend estimation and estimated the correlation function parametrically for adjusting the bandwidth selection criterion. Several methods have been proposed that utilize completely nonparametric approaches including [7, 13, 16, 20]. [8, 12, 13] studied the same problem with long-range dependent errors.

Exponential smoothing is a common means of forecasting a future realization of a time series. [10] pointed out that exponential smoothing can be viewed as a special type of nonparametric regression procedure where fitting a particular point uses only data to the left of that location. They also showed that the most common adaptive choice of the smoothing factor in exponential smoothing is identical with the cross-validation technique from nonparametric regression.

While classical kernel methods seek to find the "optimal" smoothing parameter, SiZer is based on scale-space ideas from computer vision, see [17]. Scale-space is a family of kernel smooths indexed by the bandwidth $h$. SiZer considers a wide range of bandwidths which avoids the classical problem of bandwidth selection. Furthermore, the target of a SiZer analysis is shifted from finding features in the "true underlying curve" to inferences about the "smoothed version of the underlying curve". The idea is that this approach uses all the information that is available in the data at each given scale.

Wavelet-based methods are similar to SiZer in the sense that they also look at the data at different locations and scales simultaneously. Specifically, while SiZer tries to find local minima and maxima, a wavelet analysis is usually used to achieve different goals, *e.g.* reconstructing a function or detecting change points or structural breaks. It is hard to determine the significance of trends for low frequency behavior using a wavelet analysis. In addition, it needs to control the vanishing moments parameter. Thus, smooth slow deterministic trends are difficult to be detect if one uses wavelets with a high number of zero moments because the smooth trends will be ignored by the high vanishing moments. On the contrary, SiZer does not suffer from this problem and can detect both slow and sharp trends by studying the derivatives of the smooths. Another advantage of SiZer analysis is that the inference is summarized with visualization. One can easily recognize whether a trend is significantly going up (down) or not from the family of smooths and the SiZer map. Its simplicity, interpretability and direct connection to the scale-space ideas from computer vision make SiZer attractive in general contexts.

[21] developed a dependent SiZer which conducts a goodness-of-fit test for time series. The dependent SiZer uses a true autocovariance function of an assumed model instead of estimating it from the observed data. By doing so, a goodness-of-fit test can be conducted and we can see how different the behavior of the data is from that of the assumed model. The method proposed in this paper does not need to specify an assumed model and estimates the covariance structure from the data.

The following subsections describe the local linear fit and the conventional SiZer method.

### 2.1. The local linear fit

One interesting approach to smoothing is local polynomial modeling (see e.g. [9], for a detailed overview of this method). Here are the main ideas in the time series context for trend estimation.

The data are collected at discrete points at $t = t_i$ for $i = 1, \ldots, n$, and since trend estimation in time series can be viewed as a regression problem for fixed and equispaced design with correlated errors, we can set $t_i = i$. Given the data $\{(i, Y_i), i = 1, ..., n\}$, the regression setting is

$$Y_i = m(i) + \sigma \epsilon_i, \qquad i = 1, ..., n \qquad (1)$$

where $m$ is assumed to be a smooth function and the error is assumed to be a zero mean weakly stationary process, i.e. $E(\epsilon_i) = 0$, $V(\epsilon_i) = 1$, for all $i = 1, \ldots, n$, and

$$Cov(\epsilon_i, \epsilon_j) = \gamma(|i - j|),$$

for all $i, j = 1, \ldots, n$.

Of interest is the estimation of the regression function $m$ and its derivatives at $i_0$, i.e. $m(i_0) = E(Y|i = i_0)$ and $m'(i_0)$, $m''(i_0)$, ..., $m^{(p)}(i_0)$.

The flexibility of smoothing techniques comes from only assuming smoothness of the underlying curve. This means, intuitively, that data information about the value of the regression function $m$ at $i_0$ is given by the observations at $i$ near $i_0$ and therefore these observations can be used to construct an estimator of $m(i_0)$.

By assuming that the $(p+1)^{th}$ derivative of $m(i)$ at the point $i_0$ exists, the local polynomial approach approximates the regression function $m(i)$ locally by a polynomial of order $p$. This polynomial is then fitted locally by solving a weighted least squares regression problem. In particular, in the local linear fit the function $m(i)$ is approximated by Taylor expansion of order 1 for $x$ in the neighborhood of $x_0$. The problem to be solved is then

$$\min_{\beta} \sum_{i=1}^{n} [Y_i - (\beta_0 + \beta_1(i - i_0))]^2 K_h(i - i_0)$$

where $\boldsymbol{\beta} = (\beta_0\ \beta_1)'$, $h$ is the bandwidth controlling the size of the local neighborhood and $K_h(\cdot) = \frac{1}{h} K(\frac{\cdot}{h})$, where $K$ is a kernel function, often taken to be a symmetric probability density, assigning weights to each datum point.

By Taylor expansion $\beta_0 = m(i_0)$ and $\beta_1 = m'(i_0)$, so the solution to this problem gives estimates of the regression function and its first derivative at $i_0$. More specifically,

$$\hat{\boldsymbol{\beta}} = (X^T W X)^{-1} X^T W Y$$

where $Y = (Y_1, \ldots, Y_n)^T$, the design matrix of the local linear fit at $i_0$ is

$$X = \begin{pmatrix} 1 & (1 - i_0) \\ 1 & (2 - i_0) \\ \vdots & \vdots \\ 1 & (n - i_0) \end{pmatrix}$$

and $W = diag\{K_h(i - i_0)\}$.

Because of the excellent interpretability properties of kernel type smoothers and because of their simplicity, [19] recommend the local linear fit to construct the family plot. Moreover, for a better visualization of the family of smooths, it is suggested to use a 'very wide range' of $h$ values in the log scale and for symmetry an odd number of curves should be chosen (for details see [19]).

### 2.2. The SiZer approach

To understand the development of SiZer for time series let us briefly illustrate the mathematical aspects of the original version of SiZer for regression function estimation in the specific case of fixed and equispaced design but where errors are

i.i.d.. As previously stated, the SiZer approach is based on confidence intervals for the derivative of the smoothed underlying function. These are of the form

$$\hat{m}'_h(i) \pm q(h)\hat{sd}(\hat{m}'_h(i))$$

where $q(h)$ is an appropriate Gaussian quantile. There are two critical points of interpretation.

The first point is that from the scale space point of view, the target of the SiZer analysis is shifted from the 'true underlying curve' to 'smoothed versions of the underlying curve'. In particular, instead of seeking confidence intervals for $m'(x)$, we seek confidence intervals for the scale-space version $m'_h(x) \equiv E\hat{m}'_h(x)$. In this way, the center point of each confidence interval is automatically 'correct', i.e. the interval is unbiased. This makes sense because it reflects the part of the regression curve that is available from the data at the level of resolution, $h$. In other words, for each value of the bandwidth, all the information available in the data set is considered in the inference process. For more details see [5].

The second point is that the confidence intervals are constructed in a simultaneous way. The approach is, for each level of resolution $h$, to approximate the full simultaneous confidence limit problem by $l$ independent confidence interval problems. The quantity $l$, which reflects the number of 'independent blocks', is estimated through the quantity *Effective Sample Size* ($ESS$), i.e. for the bandwidth $h$ and at $i_0$

$$ESS(i_0, h) = \frac{\sum_{i=1}^n K_h(i_0 - i)}{K_h(0)}.$$

Note that when using the uniform kernel, $ESS(i_0, h)$ is equal to the number of data points in the kernel window centered at $i_0$ with the bandwidth $h$. The number of independent blocks is then approximated by the quotient:

$$l(h) = \frac{n}{ESS(i_0, h)}. \tag{2}$$

This results in the quantile

$$q(h) = \Phi^{-1}\Big(\frac{1 + (1-\alpha)^{\frac{1}{l(h)}}}{2}\Big). \tag{3}$$

where $\alpha$ is a significance level.

$ESS$ is also considered to decide when the normal approximation is inadequate, i.e. for $ESS < 5$ no conclusion can be drawn. In particular, the black, white, and intermediate grey are used on the scale space set

$$\{(i_0, h) : ESS(i_0, h) \geq 5\}. \tag{4}$$

Again detailed discussion of SiZer is available in [5].

## 3. SiZer for time series

For time series trend estimation, i.e. in the context of a regression problem with fixed and equispaced design with correlated errors, SiZer can be developed by modeling the error structure and by adjusting the confidence limits for the derivative of the smoothed underlying function according to the assumed error structure.

As suggested by [19], the family of smooths is constructed by considering a 'very wide range' of bandwidths $h$ in the log scale and, in particular, the number of curves is here taken to be 11.

### 3.1. The variance

For correlated data, the variance of the local polynomial estimator is given by

$$V(\hat{\beta}|X) = (X^TWX)^{-1}(X^T\Sigma X)(X^TWX)^{-1} \qquad (5)$$

where, for the assumed correlation structure, $\Sigma$ is the kernel weighted covariance matrix of the errors where the generic element is given by

$$\sigma_{ij} = \gamma(|i-j|)K_h(i-i_0)K_h(j-i_0). \qquad (6)$$

A sensible estimate of the variance (5) is based on estimating $\gamma$ in (6), by the sample autocovariance function of the observed residuals from a 'pilot smooth', using the pilot bandwidth $h_p$. One could take $h_p = h$, but this leads to a confounding of the different notions of 'scale' and 'dependence structure'. In other words, for dependent errors, the estimate of $\gamma$ depends on the choice of $h_p$, unlike independent errors. A small $h_p$ assumes i.i.d. or weakly correlated errors, and a large one corresponds to strongly correlated errors. Thus, if one takes $h_p = h$, those features which appear in a particular row of a SiZer map might not be clearly interpreted as significant trends or a wrong assumption on dependence structure. Hence, we treat $h$ and $h_p$ separately, which means that in the dependent case, another dimension needs to be added to the SiZer plot. We approach this via a series of SiZer plots, indexed by the pilot bandwidth $h_p$, which represent the different trade-offs available between trend and dependence. This is the key to our visualization, which is further developed in Section 4.

### 3.2. The quantile

For positively autocorrelated errors, the family of smooths typically varies substantially as the bandwidth is varied. The reason is that the strong sporadic patterns that characterize such errors behave similarly to high frequency regression components, which appear in the smooth for a wide range of bandwidths. On the other hand, for negatively autocorrelated errors, the tendency of data points to alternate above and below the regression function gives a family of smooths which changes less as a function of the bandwidth. This is one way of seeing that the amount of information, about the underlying smooth regression function, that is available in i.i.d. data, is not the same as the amount of information available in correlated data. Positively correlated data contain 'less information' about the regression function than i.i.d. data, while negatively correlated data contain 'more information' about the regression function than i.i.d. data. Using statistical information ideas, a simple measure of 'information in the data', on the scale of sample size, is provided by the ratio

$$n^\star = \frac{\sigma^2}{Var(\bar{Y})},$$

where

$$Var(\bar{Y}) = \frac{\sigma^2}{n} + \frac{2}{n}\sum_{k=1}^{n-1}\left(1 - \frac{k}{n}\right)\gamma(k).$$

The ratio of $n^\star$ to $n$ gives a version of the $ESS$, i.e. the *Effective Sample Size*, which properly reflects the type and the magnitude of the correlation structure to give correct simultaneous inference:

$$ESS^\star(i, h) = \frac{n^\star}{n} \frac{\sum_{i=1}^{n} K_h(i - i_0)}{K_h(0)}.$$

For independent data, $n^\star = n$, so $ESS^\star = ESS$ from conventional SiZer. But for correlated data, $n^\star$ is smaller or larger than $n$ depending on the type of correlation, i.e. observations have less effect if they are positively correlated (since they are 'less informative' than i.i.d. observations) and more effect if they are negatively correlated (since they are 'more informative' than i.i.d. observations).

The computation of $l$ in (2) and $q$ in (3) remains unchanged.

When the data are positively autocorrelated, the number of independent problems is larger than in the i.i.d. case, so the resulting confidence intervals are longer. Longer intervals are more likely to contain the value 0, so fewer features are flagged as statistically significant. On the other hand, for negatively correlated data (a situation that is rare in real data but is worth considering), SiZer for time series can detect those features that could be hidden in the family of smooths by the alternating pattern of data points above and below the regression function.

In the long-range dependent case, the effective sample size becomes much smaller than that of the i.i.d. case. Since the time series is strongly positively correlated, $n^\star$ will be much smaller than $n$, which results in the smaller $ESS^\star$. As a result, there will be more chances for the effective sample size $ESS^\star$ of each pixel to be less than 5, which produces more darker grey regions in a SiZer map (no test being made).

Finally the quantity $n^\star$ has to be estimated (since $\sigma^2$ and $\gamma$ are unknown). The estimate given by the sample autocovariance function cannot be expected to give good results since it is inconsistent. A simple approach to this problem is to divide the sample into $p = \sqrt{n}$ groups (so as to have a reasonably large number of groups with a reasonably large number of observations), and to estimate the variance of $\bar{Y}$ by

$$\widehat{Var}(\bar{Y}) = \frac{1}{p}\frac{1}{p-1}\sum_{j=1}^{p}(\bar{Y}_j - \bar{\bar{Y}})^2$$

where $\bar{Y}_j$ is the mean in the group $j$ and where

$$\bar{\bar{Y}} = \frac{1}{p}\sum_{j=1}^{p}\bar{Y}_j.$$

## 4. Simulated examples

In this section, we will illustrate some examples of SiZer for time series, chosen from many simulated data sets. In Section 4.1, we will illustrate the development of the methodology and graphical presentation of the results. Section 4.2 addresses a coverage probability of SiZer.

### *4.1. Illustration of SiZer*

The simulation study has been carried out by considering different combinations of trend, error structure and noise level. A series of different data sets have been simulated from the model (1) for different choices of $m$, $\{\epsilon_i\}$ and $\sigma$, and for different values of $n$.

The trends that were considered are characterized by a different number of 'peaks and valleys', differently located and with different amplitudes. For each chosen trend we have considered i.i.d. errors and positively and negatively correlated errors. In particular, i.i.d. errors have been simulated from $N(0,1)$ while correlated errors have been simulated from an autoregressive process of order 2 with high correlation at lag one (for the examples of the simulation study shown in this paper, the value of the autocorrelation function at lag one for positively correlated errors is 0.97 while for negatively correlated errors the value of the autocorrelation function at lag one is $-0.89$). Moreover, in each case, three different noise levels were chosen to represent 'low', 'medium' and 'high' variability (in the examples shown below, these values were respectively 1, 20 and 50).

While we do not report all the results in this paper, several interesting cases among the models considered had the trend:

$$m(i) = i + 10sin\left(\frac{i}{40}2\pi\right) \qquad i = 1, ..., n. \qquad (7)$$

Given the model in (1) and the trend in (7), let us consider a simulated time series, that we will indicate by sm1, where errors are positively correlated and a 'medium' level of variability is chosen, and $n = 200$. An important issue is how many SiZer plots, should be shown. There will be a different plot for each pilot bandwidth $h_p$, showing different trade-offs between trend and serial correlation. If $h_p$ takes all 11 values of the bandwidth $h$ then the complete series of SiZer plots will be 11 in number. The family plot and the complete series of 11 SiZer plots for the simulated time series sm1 is given in Figure 3.

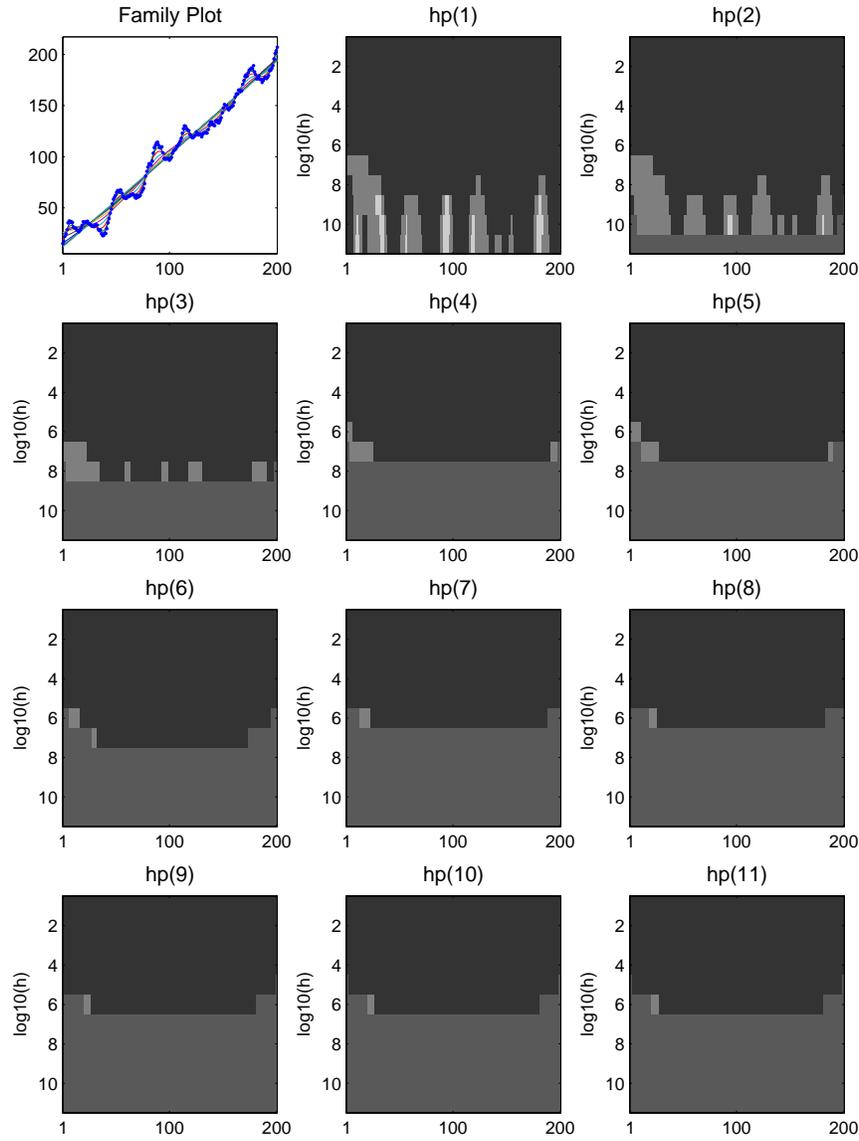

Fig 3. *The family plot and the complete series of 11 SiZer plots, indexed by $h_p$, for the simulated time series sm1.*

The family plot shows a family of smooths with a strong upward trend. The first SiZer plot (titled $hp(1)$) has very small $h_p$. Thus, the error structure used to construct this plot is estimated from the residuals of the closest curve of the family of smooths to the data points, which is essentially i.i.d.. From this plot we observe some significant structure at the finest levels of resolution, i.e. for the smallest values of the bandwidth. As the level of resolution decreases, less and less structure is significant until the curves are significantly increasing everywhere.

In the second SiZer plot (titled $hp(2)$), which represents a situation of slightly correlated errors, less structure than in the previous plot, as expected, appears as significant. Moreover, at the finest level of resolution the data are too sparse, in terms of $ESS^\star$, to draw any conclusion, i.e. the bottom line is now shaded darker grey, as discussed in Section 2.2.

As we move from the third to the last SiZer plot (i.e. from the plot titled $hp(3)$ to the plot titled $hp(11)$), an increasing amount of correlation appears in the error component, so that only the linear upward trend seems to be significant at every level of resolution, where conclusions can be drawn. Also, at the finest levels of resolution there is less perceived useful information in the data, which means more data sparsity, so more bottom lines of the SiZer plots are shaded darker grey.

Statistically significant structure in a curve can be hidden by a strong linear component. When this happens, as in this example, it can be useful to detrend the time series before using SiZer. Figure 4 shows the complete series of SiZer plots, for the same data set after linear detrending (indicated by sm1d).

Now, many features appear in the family of smooths which is no longer closely following an increasing line. In particular, in the first SiZer plot, many peaks and valleys are significant for most of the levels of resolution. As the errors include increasing correlation, these features are significant but with a less precise location and for a smaller number of levels of resolution. From the seventh plot to the last, all of the structure in the data is explained by the error component and no significant structure is highlighted in the trend at any level of resolution.

This example clearly shows that the complete series of SiZer plots is too long. The simultaneous view of all 11 SiZer plots is hard to comprehend and the information contained in several such plots is often redundant. This motivates the choice of a subset of SiZer plots.

We develop a method for effective choice of a subset of these, which usually gives good representatives of the major different correlation structures. We found 4 plots usually conveyed the needed information. Our choice among the 11 plots is intended to reflect 'a wide array of trade-offs between trend and dependence'. A simple numerical measure of this trade-off is the $IR$ (Indicator of the Residual component), which takes values from 0 to 1:

$$IR(h_p) = \frac{\sum_{i=1}^n e_{h_p,i}^2}{\max_{h_p} \sum_{i=1}^n e_{h_p,i}^2}, \qquad (8)$$

where $e_{h_p,\cdot}$ are the residuals obtained from the pilot bandwidth $h_p$. When the pilot bandwidth $h_p$ is large, the 'dependence component of the data' appears strongly in the residuals, i.e. is viewed as 'noise', and *IR* is close to 1. On the other hand, when the pilot bandwidth $h_p$ is small, the 'dependence component of the data' appears strongly in the pilot smooth, i.e. is viewed as 'trend', and the *IR* is close to 0. Intermediate values of *IR* reflect intermediate trade-offs.

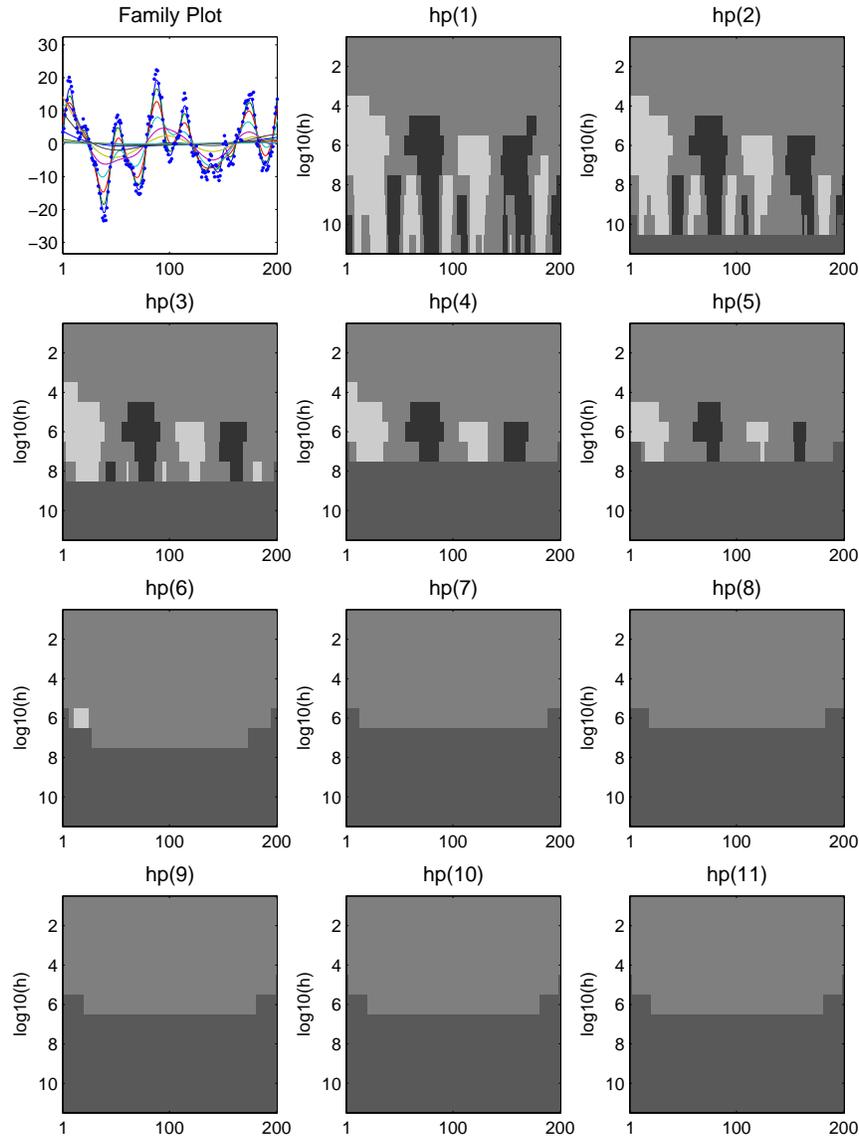

Fig 4. *The family plot and the complete series of 11 SiZer plots, indexed by $h_p$, for the simulated and linearly detrended time series sm1d.*

A good reflection of the range of trade-offs came essentially from using the SiZer plots chosen (from the set of 11) for which $IR$ is closest to 0%, 25%, 50% and 75%. This implies that the first plot to be chosen should always be the first of the 11 plots. However, since the error structure used to construct this plot is estimated from the residuals given by a smooth for which the degree of overfitting may be too high (mostly when data are positively correlated), a second option is considered. When this is the case, the second of the 11 plots is chosen as the closest to 0%, i.e. as 'more representative of the independent case'. And if this plot corresponds to a value of $IR$ which is also the closest to 25%, the 3 remaining plots will automatically be chosen by considering 25%, 50% and 75% of the difference between the values of $IR$ for the 11th and the 2nd plots of the complete series.

When only 4 of the SiZer plots are displayed, it is useful to stay in touch with the trend-dependence trade-off that each represents. This is done by several graphical devices. First, all 11 values of $IR$ that are considered are displayed as a bar graph, and the chosen 4 are highlighted. Second, for each chosen trade-off, we add plots showing the pilot smooth with bandwidth $h_p$, and the residuals from that smooth. The pilot smooths show which component of the data are viewed as trend, in that particular trade-off. The residuals give a visual impression of the component of the data used in the covariance estimate.

This is the visualization chosen for SiZer for time series. Figure 5 shows this graphical device for sm1d.

The data are shown in the first plot above on the left (the continuous line shows $m(i)$, the deterministic part of the simulated time series), while the next graphic on the right is the family plot. Further right, is the bar diagram previously discussed. The second and the third series of plots represent, respectively, the smooths and the residuals.

Note that the chosen SiZer plots give a representative sample of the complete series of SiZer plots shown in Figure 4. No relevant information is lost and the graphical representations of the smooths and the residuals associated to each SiZer give useful insights about each trend-dependence trade-off. This demonstrates the power of the SiZer method for investigating trends in time series: one graphical presentation displays all of the relevant information about statistical significance.

### *4.2. Type I error and power of SiZer plot*

In this section, a type I error and a power of multiple tests in a SiZer map are calculated by simulating three different types of noise. These probabilities will provide how often the SiZer plots are in agreement with the 'truth' in presence of noises. The noises considered here are white noise (i.i.d.) with $\sigma = 1$, MA(1) (weakly correlated) with the coefficient 0.9 and $\sigma = 1$, and fractional Gaussian noise (strongly correlated) with the Hurst exponent $H = 0.9$ and $\sigma = \sqrt{20}$. The trend is added to these three noises is

$$m(i) = \sin(6\pi i/n) - i/n, \quad i = 1, \cdots, n \qquad (9)$$

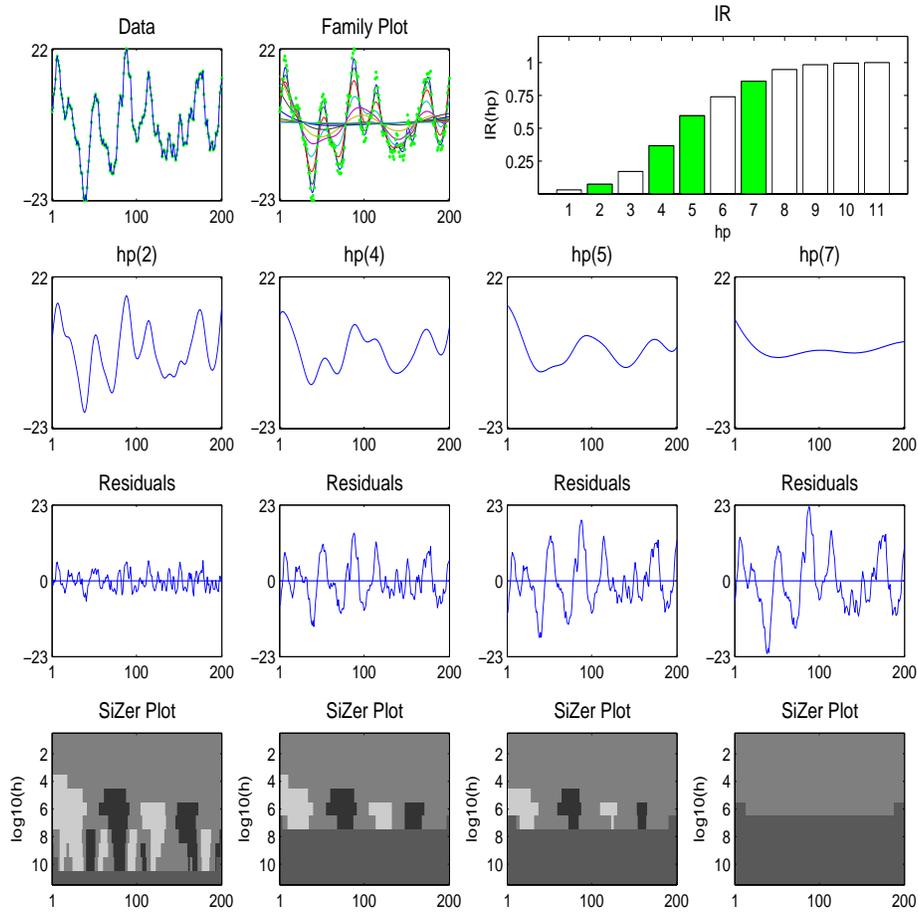

FIG 5. *SiZer for time series for the simulated and linearly detrended time series sm1d.*

where $n = 400$. Thus, the trend has a sine wave and a decreasing trend as depicted in the upper leftmost of Figure 6. The upper middle plot shows the family smooths of the simulated MA(1).

The type I error and the power are calculated as follows. First, the SiZer map of true derivative is created by plugging $m$ in (9) into SiZer. Since the estimation of an autocovariance function is not needed, only one SiZer map is produced, and the second row of Figure 6 shows the true derivative map (the plot is repeated four times at each column for easy comparison with the third row) for (9). Second, SiZer maps are created by plugging $Y_i$ ((9) with noise MA(1)) into SiZer and four of them are selected by $IR$ statistics in (8), which are shown in the third row of Figure 6. Then, each SiZer map is compared with the true derivative map pixel by pixel over the set (4). Here, we define the Type

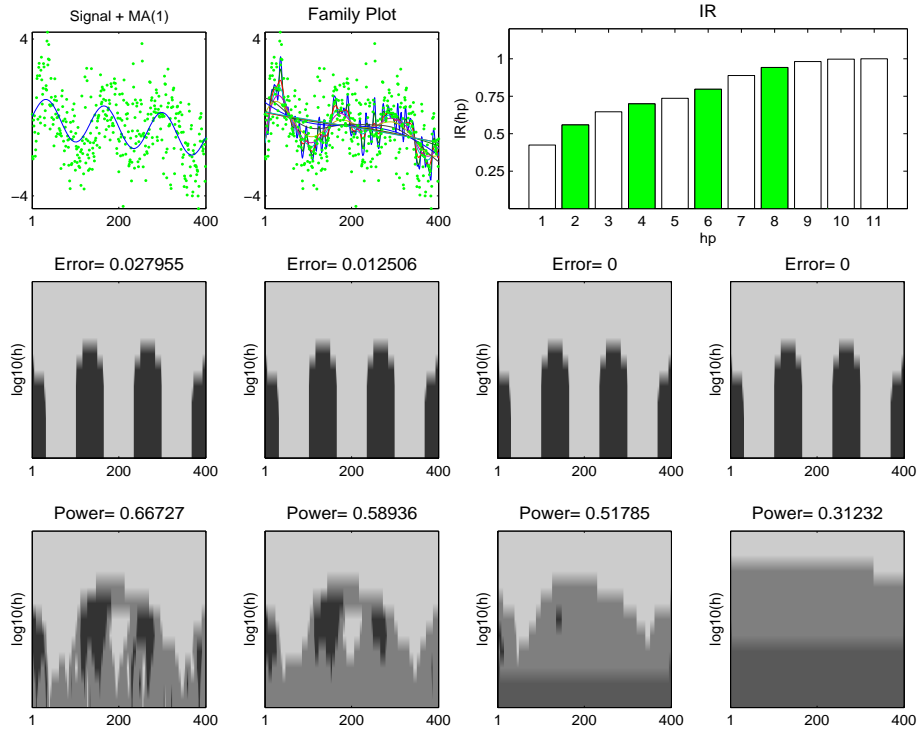

Fig 6. *SiZer for the simulated time series examples with type I error and power.*

I error and the power as follows. The type I error counts pixels with (true = positive, SiZer map = negative) or (true = negative, SiZer map = positive) or (true=zero, SiZer map = positive or negative). The final error probability is obtained by dividing these counts by total pixels in (4). The power is defined as 1−(# of (true = positive or negative, SiZer map = zero)/ (total pixels in (4))). This can be considered as the probability of decision undeferred. In Figure 6, the titles of the second and third rows are examples of the type I error and the power.

Table 1 shows the type I error and the power for the three simulated examples. In the table, the $hp1$ means the first bandwidth selected by $IR$, and the $hp2$ the second selected by $IR$, and so on. Simulations are repeated 100 times and the table reports their mean, median and maximum.

For the white noise and MA(1) cases, the maximum of Type I error is 0.0459 throughout all the bandwidths and the 100 repetitions, which is very low. This implies that SiZer analysis provides high coverage probabilities which are close to 1 and makes few mistakes in its decision for weakly correlated errors. However, for the fractional Gaussian case, the mean and the median of Type I error is above 0.3. Also, considering the difference between the maximum and the minimum, the variation is also large throughout the repetitions. This suggests that SiZer needs a better variance estimator for strongly correlated errors such as fractional Gaussian noise. We leave this for future work. As the pilot bandwidth increases, the type I error decreases, but the power also decreases because the SiZer map cannot find the detailed features because of the coarse resolutions.

TABLE 1
*Type I error and Power*

|  |  | Type I |  | Error |  |  | Power |  |  |
|---|---|---|---|---|---|---|---|---|---|
| Noise |  | Mean | Med | Max | Min | Mean | Med | Max | Min |
| $N(0,1)$ | $hp1$ | 0.0163 | 0.0150 | 0.0307 | 0 | 0.7627 | 0.7630 | 0.8059 | 0.7255 |
|  | $hp2$ | 0.0001 | 0 | 0.0038 | 0 | 0.5636 | 0.5632 | 0.6695 | 0.4681 |
|  | $hp3$ | 0 | 0 | 0 | 0 | 0.4993 | 0.5061 | 0.5749 | 0.3802 |
|  | $hp4$ | 0 | 0 | 0 | 0 | 0.4666 | 0.4762 | 0.5305 | 0.3428 |
| MA(1) | $hp1$ | 0.0260 | 0.0260 | 0.0459 | 0.0116 | 0.7003 | 0.6993 | 0.7595 | 0.6552 |
|  | $hp2$ | 0.0089 | 0.0075 | 0.0298 | 0 | 0.6219 | 0.6214 | 0.6889 | 0.5246 |
|  | $hp3$ | 0.0008 | 0 | 0.0139 | 0 | 0.5143 | 0.5155 | 0.6571 | 0.2240 |
|  | $hp4$ | 0.0001 | 0 | 0.0095 | 0 | 0.4351 | 0.4517 | 0.5513 | 0 |
| FGN | $hp1$ | 0.3310 | 0.3589 | 0.5759 | 0.0550 | 0.6227 | 0.6516 | 0.7380 | 0.3255 |
|  | $hp2$ | 0.2718 | 0.2160 | 0.5425 | 0.0366 | 0.5054 | 0.5386 | 0.6566 | 0.1264 |
|  | $hp3$ | 0.2275 | 0.1539 | 0.5706 | 0 | 0.4208 | 0.4777 | 0.6317 | 0.0280 |
|  | $hp4$ | 0.1600 | 0.0405 | 0.6186 | 0 | 0.2904 | 0.3834 | 0.6533 | 0 |

## 5. Real examples

The real data sets shown here are the Deaths data set, i.e. the monthly number of accidental deaths in US from 1973 to 1978 (thousands) and the Chocolate data set, that is considered in the introduction of this paper. As the Chocolate data set, the Deaths data set comes with the software companion to [4].

### *5.1. Deaths data*

Figure 7 shows SiZer for time series for the Deaths data set, after deseasonalizing and linear detrending.

Assuming 'i.i.d.' errors (first SiZer plot on the left) only some features of the family of smooths are significant at the finest levels of resolution. The strongest feature is the minimum around $i = 36$ for intermediate values of the bandwidth. But for smaller bandwidths, there is a significant increase near $i = 20$. For the largest value of the bandwidth the curve is neither significantly increasing nor decreasing.

For slightly correlated errors (second SiZer plot on the left), the major minimum again turns out to be significant but for a smaller range of bandwidths than in the previous plot (for the smallest values of the bandwidth no conclusion can be drawn and for the largest values no feature is significant). When the correlation increases, the number of values of the bandwidth for which no conclusion can be drawn increases while the minimum previously highlighted is still significant but only for a few values of the bandwidth and with a much less precise location. For a higher degree of correlation, no feature turns out to be significant, as we can see from the graphical presentation of the associated residuals and smooth. For such correlation structure every feature can be explained by the error component, thus not resulting in significant trend.

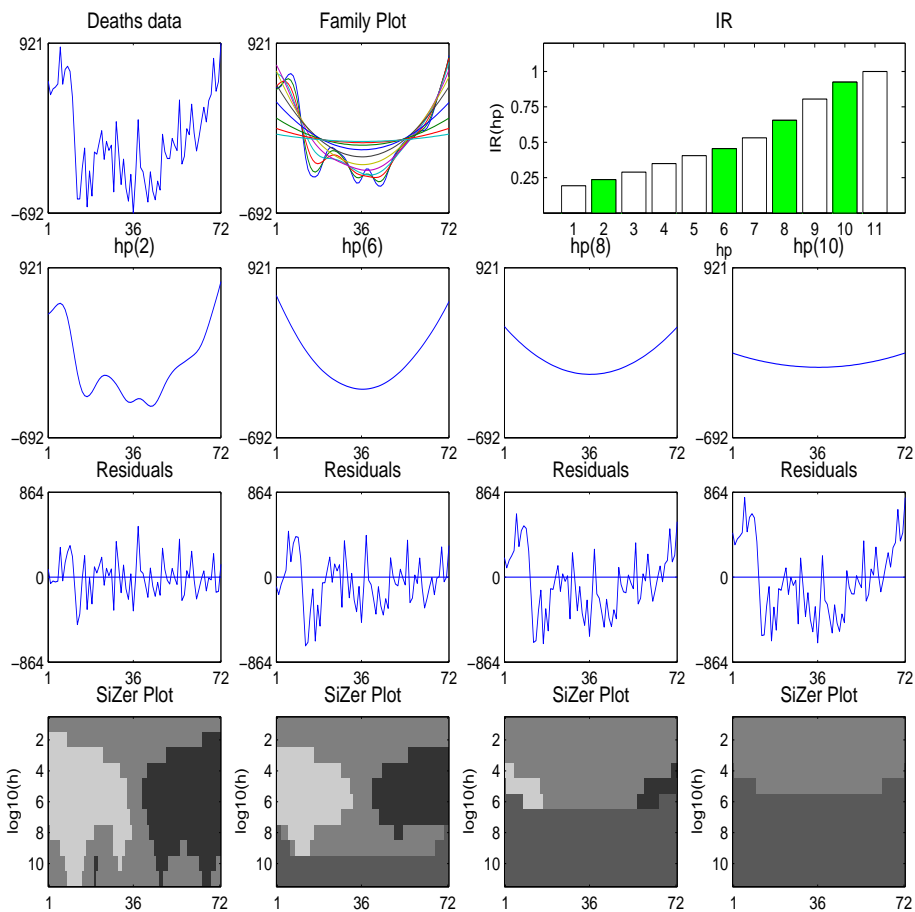

Fig 7. *SiZer for time series for the Deaths data set.*

In this example, according to SiZer for time series, the only feature that appears to be significant at most of the levels of resolution, if we assume that errors are not strongly correlated, is the 'valley' around the third year of observation. For larger error correlation there is no significant trend.

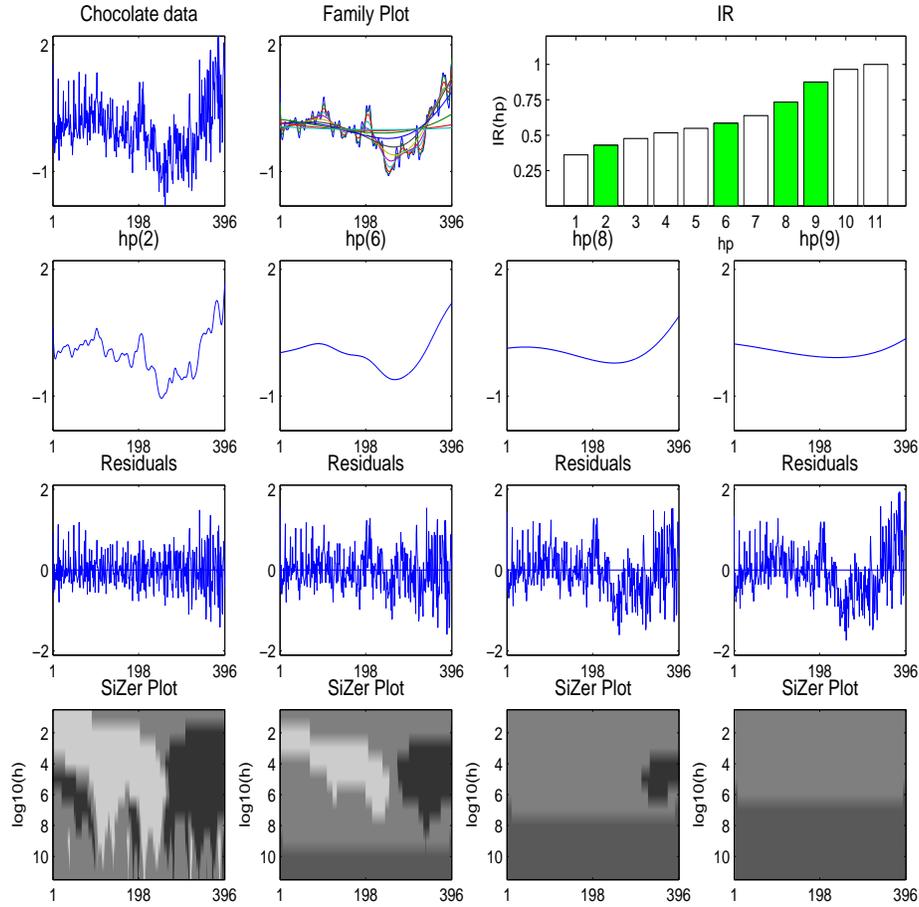

Fig 8. *SiZer for time series for the Chocolate data set.*

### 5.2. Chocolate data

Figure 8 shows SiZer for time series for the Chocolate data set. As in the previous example, this data set has been deseasonalized and linearly detrended. But here the strong upward trend, driven by increasing chocolate production due to increasing population size, seems to hide much more structure than in the previous case, as we can see once we detrend the data, especially when the errors are assumed to be 'i.i.d.'. Nevertheless, SiZer for time series still gives useful insights about the trends under study, using the same 4 choices of SiZer plots from the complete series of 11.

For 'i.i.d.' errors (first SiZer plot on the left) many 'peaks and valleys' are significant at the finest levels of resolution, i.e. for the smallest values of the bandwidth. As the level of resolution decreases, fewer and fewer features appear to be significant.

For slightly correlated errors, no conclusion can be drawn at the finest levels of resolution while for intermediate values of the bandwidth the major minimum around $i = 250$ (which corresponds to year 1978) is the only significant feature. For the highest value of the bandwidth the curve is neither significantly increasing nor decreasing.

As the correlation increases no significant structure is highlighted with the only exception being some intermediate values of the bandwidth where the curves are significantly increasing on the right end of the time domain.

For the strongly correlated error assumption, no significant structure appears at any level of resolution.